\documentclass[%
 reprint,
amsmath,amssymb,aps,
]{revtex4-2}

\usepackage{graphicx}
\usepackage{dcolumn}%
\usepackage{bm}
\usepackage{hyperref}

\begin{document}
\title{On the origin of CP symmetry violations}
\author{J-M Rax}
\email{jean-marcel.rax@universite-paris-saclay.fr}
\affiliation{Universit\'{e} de Paris-Saclay\\
IJCLab-Facult\'{e} des Sciences d'Orsay\\
91405 Orsay France}
\date{\today}

\begin{abstract}
Experiments devoted to charge parity (CP) violation 
are normally interpreted 
by adjusting the
elements of the Cabibbo-Kobayashi-Maskawa matrix to the
measured violation parameters. However, the physical origin of these violations 
remains an open issue. To resolve this issue, 
the impact of Earth's gravity on meson oscillations 
is analysed. 
The effect of gravity is to couple flavour oscillations 
to quark zitterbewegung oscillations, and this coupling induces 
a superposition of CP eigenstates.
The three types of CP violation effects result
from this gravity-induced mixing. 
The three associated violation parameters are predicted in agreement 
with experimental data. The
amplitude of the violation is linear with respect to gravity,
so this new mechanism allows us to envisage cosmological evolutions 
that provide the observed baryonic asymmetry of the universe.
\end{abstract}

\maketitle

\section{Introduction}
Following the first observations of long-lived kaons anomalous decays \cite{1}, 
several observables
associated with flavored neutral mesons Charge-Parity violation (CPV) have been
identified, measured and interpreted during the past decades. Neutral mesons
experiments dedicated to CPV are interpreted within the framework of the
standard model (SM) through the adjustment between ({\it i}) the
Kobayashi-Maskawa (KM) complex phase \cite{2}, in the
Cabibbo-Kobayashi-Maskawa (CKM) matrix \cite{2,3}, and ({\it ii}) the
measured violation parameters.
Despite the success of the SM-CKM interpretation based on the adjustment 
between the CKM matrix and the experimental data, the physical origin of CPV remains an
open issue. To resolve this issue, we demonstrate that {\it gravity induced
CPV} (GICPV) provides a pertinent and accurate framework to interpret the
most documented experimental evidences of CPV and to predict the violation
parameters in agreement with the experimental data. This set of new results 
does not rely on the adjustment of free parameters. 

As a consequence of the accuracy of GICPV
quantitative predictions, we conjecture that far from any massive object,
i.e. in a flat Lorentzian space-time, the CKM matrix must be free from any
CPV phase: CPV effects appears to be gravity induced near massive objects
like Earth.

The three main types ({\it i}, {\it ii} and {\it iii}) of CPV experiments
are analyzed here: ({\it i}) indirect CPV in the mixing observed with
neutral kaons $K^{0}/\overline{K}^{0}$, the observable of these experiments
is the parameter $\mathop{\rm Re}\varepsilon $ \cite{4}; ({\it ii}) direct
CPV in decays into one final state, the observable of these experiments is
the parameter $\mathop{\rm Re}\left( \varepsilon ^{\prime }/\varepsilon
\right) $ \cite{5}; ({\it iii}) CPV in interference between decays with and
without mixing, the observable of these $B^{0}/\overline{B}^{0}$ experiments
is the angle $\beta $ \cite{5}. A forth ({\it iv}) experimental evidence of
CPV is to be considered: the observed dominance of baryons over antibaryons
in our universe because CPV is one of the necessary condition to build
cosmological evolution models compatible with the baryon-antibaryon 
abundance asymmetry \cite{6}.

The SM provides a framework to interpret Earth based CPV experiments such as
({\it i}, {\it ii} and {\it iii}), but this interpretation,
incorporated into cosmological evolution models, fails, by several orders of
magnitude, to account for this ({\it iv}) major CPV evidence. To explain how
our matter-dominated universe emerged during its early evolution we need to
identify a CPV mechanism far larger than the KM one. Beside its potential to
predict accurately the measured parameters ($\varepsilon $, $\varepsilon
^{\prime }$, $\beta $) associated with types ({\it i}, {\it ii} and {\it iii}%
) CPV experiments on Earth, the new GICPV mechanism opens very interesting
perspectives to set up cosmological models displaying an asymmetric
baryogenesis compatible with the present state of our universe. Indeed,
during the early stages of the cosmological evolution,  gravity/curvature was
far more larger than on Earth today and GICPV, which is a linear function of
the gravitational field, opens an avenue to resolve the present
contradiction between the very small SM-KM CPV and the very large CPV
needed to build a pertinent model of our matter-dominated universe.

In this paper we demonstrate that the small coupling, induced by Earth's
gravity, between ({\it i}) fast {\it quarks zitterbewegung oscillations,} at
the velocity of light, inside the mesons and ({\it ii}) {\it strangeness
oscillations} $\Delta S=2$, or {\it bottomness oscillations} $\Delta
B^{\prime }=2$, provides both a qualitative explanation of CPV and
quantitative predictions of the CPV parameters $\varepsilon $, $\varepsilon
^{\prime }$ and $\beta $ in agreement with the most recent experimental
measurements reviewed in {\it PDG 2024} \cite{5}.

The fact that the combination of gravity and quark zitterbewegung 
is a plausible candidate 
to explain
the origin of CPV can be understood heuristically as follows. 
This heuristic argument
is restricted to the $K^{0}/\overline{K}^{0}$ case and it
can be easily extended to $B^{0}/\overline{B}^{0}$.

The Hamiltonian describing $\Delta S=2$ kaons oscillations in a flat
Lorentzian space-time, far from any massive objects, is $\widehat{H}$. The
observed meson eigenstates, $\left| M_{1}\right\rangle $ and $\left|
M_{2}\right\rangle $, are solutions of 
$
\widehat{H}\cdot \left| M_{1/2}\right\rangle =E_{1/2}\left|
M_{1/2}\right\rangle   \label{c12}
$
where, in the meson rest frame, the eigenenergy, $E_{1}$ and $E_{2}$, are
the inertial masses: $m+
\delta m/2$ and $m-
\delta m/2$. The Hamiltonian $\widehat{H}$
commute with the CP operator $\left[ \widehat{H},\widehat{CP}\right] =0$,
thus these energy eigenstates are also CP eigenstates $\widehat{CP}\left|
M_{1/2}\right\rangle =\pm \left| M_{1/2}\right\rangle $. Consider now a CPV
perturbation $\widehat{V}$ $\ll \widehat{H}$\ that might slightly mix the CP
eigenstates $\left| M_{1/2}\right\rangle $ and shift the eigenenergy $E_{1/2}
$. With such a perturbation the observed eigenstates, $\left|
M_{1}^{V}\right\rangle $ and $\left| M_{2}^{V}\right\rangle $, are solutions
of 
\begin{equation}
\left( \widehat{H}+\widehat{V}\right) \cdot \left| M_{1/2}^{V}\right\rangle
=E_{1/2}^{V}\left| M_{V1/2}\right\rangle \text{.}  \label{c1}
\end{equation}
The observed shifted eigenvalues $E_{1/2}^{V}$ and mixed eigenstates $%
M_{V1/2}$, solutions of equation (\ref{c1}), are given by the usual
first order perturbation theory:

\begin{equation}
\left| M_{1/2}^{V}\right\rangle =\left| M_{1/2}\right\rangle +\frac{%
\left\langle M_{2/1}\right| \widehat{V}\left| M_{1/2}\right\rangle }{%
E_{1/2}-E_{2/1}}\left| M_{2/1}\right\rangle \text{.}  \label{c2}
\end{equation}
On Earth, for a particle with a gravitational mass $m$, we consider the
small Newtonian energy $V\sim $ $mG_{N}M_{\oplus }/R_{\oplus }$. The matrix elements $\left\langle M_{1/2}\right| \widehat{V}\left|
M_{1/2}\right\rangle \sim mG_{N}M_{\oplus }/R_{\oplus }\sim $ $%
10^{-9}mc^{2}$ and, {\it a priori},  $\left\langle M_{2/1}\right| \widehat{V}\left|
M_{1/2}\right\rangle =0$. This perturbation induces a small energy shift of
the eigenenergy, which is not observable, and does not induces a CP
eigenstates mixing.

To pursue the analysis of the GICPV hypothesis, we consider a
 {\it CP violating vertical position fluctuation}
 $x$ and use the first term of the Taylor expansion, with
respect to $x\ll R_{\oplus }$, of \ the energy\ $mG_{N}$ $M_{\oplus }/\left(
R_{\oplus }+x\right) $. The first term of the Taylor expansion is $mgx$ with 
$g$ = $G_{N}M_{\oplus }/R_{\oplus }^{2}$ = $9.8$ m/s$^{2}$. The Compton
wavelength, $\lambda _{C}$ $=\hbar /mc\sim 4\times 10^{-16}$ m, provides an
approximate upper bound of the size of any vertical fluctuations: $x\sim \lambda
_{C}$. The expected, if any, matrix element $\left\langle M_{2/1}\right| 
\widehat{V}\left| M_{1/2}\right\rangle $ $\sim mg\lambda _{C}=\hbar g/c\sim
2.1\times 10^{-23}$ eV is rather small in front of $E_{1}-E_{2}\sim \delta
mc^{2}$ $\sim 1.7\times 10^{-6}$ eV. Thus the
expected GICPV mixing (\ref{c2}) is given by 
\begin{equation}
\left| M_{1/2}^{V}\right\rangle \sim \left| M_{1/2}\right\rangle +\frac{%
\hbar g}{\delta mc^{3}}\left| M_{2/1}\right\rangle \text{.}
\end{equation}
The order of magnitude $\hbar g/\delta mc^{3}$ $\sim 10^{-17}$ \ is far
smaller than the experimental one $\left\langle M_{2/1}\right| \widehat{V}_{%
\text{exp}}\left| M_{1/2}\right\rangle /\delta mc^{2}\sim 10^{-3}$ \cite{4}.
This leads to the conclusion that this naive point of view does not provide
an explanation to the origin of CPV. The small parameter $\hbar g/\delta
mc^{3}$ has been identified in Ref. \cite{7} by Fischbach who reaches the
same negative conclusion. Fischbach also noted that the smallness of the
parameter $\hbar g/\delta mc^{3}$ can be compensated by the large parameter $%
m/\delta m$ to provide a right order of magnitude, this puzzling remark was
later used to explore the hypothesis of antigravity as the origin of CPV 
\cite{8}.

The previous negative conclusion about GICPV can be reevaluated if we
consider the interplay between  fast quarks zitterbewegung oscillations
at the velocity of light, inside the mesons, and strangeness oscillations $%
\Delta S=2$. 

Free or bound spin 1/2 fermions, like quarks, are well known to display the so
called zitterbewegung (nonintuitive) behavior: a quiver ({\it zitter}) motion
({\it bewegung}), on a length scale given by the Compton wavelength, at an
instantaneous velocity equal to the velocity light $c$ \cite{9}. In
addition to this zitterbewegung oscillation, the strangeness oscillation $K^{0}%
\rightleftharpoons \overline{K}^{0}$ takes place at the frequency $\delta
mc^{2}/\hbar $ \cite{4}. During one particle-antiparticle transition $%
K^{0}\rightleftharpoons \overline{K}^{0}$ the quarks zitterbewegung oscillation
accumulate an energy $mgc\left( \hbar /\delta mc^{2}\right) $. The expected,
if any, matrix element $\left\langle M_{2/1}\right| \widehat{V}\left|
M_{1/2}\right\rangle \sim mg\hbar /\delta mc$ is responsible of a mixing
of the CP eigenstates 
\begin{equation}
\left| M_{1/2}^{V}\right\rangle \sim \left| M_{1/2}\right\rangle +\frac{%
m\hbar g}{\delta m^{2}c^{3}}\left| M_{2/1}\right\rangle \text{.}
\end{equation}
The numerical value $m\hbar g/\delta m^{2}c^{3}$ $\sim 10^{-3}$ leads to the
conclusion that this last point of view might provide an explanation to
the origin of CPV and so requires a deeper analysis.

The new interpretation of CPV experiments presented below is based on the
usual Hamiltonian of Lee, Oehme and Yang (LOY) \cite{10,11}, completed here
with Newtonian gravity. Neutral mesons oscillations such as $%
K^{0}\rightleftharpoons \overline{K}^{0}$ $\ $ and $B^{0}\rightleftharpoons 
\overline{B}^{0}$ are very low energy oscillations ($10^{-6}-10^{-4}$ eV), 
so there is no need to rely on quantum field theory and the usual
LOY Hamiltonian offers the pertinent framework to describe a low energy quantum
oscillation between two quantum states slightly perturbed by gravity.

The study presented below complements a previous study based on two coupled
Klein-Gordon equations describing $K^{0}/\overline{K}^{0}$ evolution on a
Schwarzschild metric \cite{12}, rather than a Newtonian framework with \ two
coupled Schr\"{o}dinger equations used here. The results given by the Newtonian model are similar to those of this
previous Einsteinian model \cite{12}, these results are thus model
independent.

This paper is organized as follows, in the next section we briefly review
the LOY Hamiltonian without CPV, then, in section 3, the experimental CPV parameters 
are defined. The impact of Earth's gravity is
considered in section 4 where, to describe neutral mesons oscillations $%
M^{0}\rightleftharpoons \overline{M}^{0}$on Earth, the CP conserving LOY
Hamiltonian, presented in section 2, is completed with a gravity term.

The study of type ({\it i}), ({\it ii}) and ({\it iii}) GICPV are developed
in sections 5, 6 and 7. We consider specifically type ({\it i}) and ({\it ii}%
) CPV for $K^{0}/\overline{K}^{0}\sim \left( d\overline{s}\right) /\left( 
\overline{d}s\right) $ and type ({\it iii}) CPV for $B^{0}/\overline{B}%
^{0}\sim \left( d\overline{b}\right) /\left( \overline{d}b\right) $. Section
8 provides a brief comment on others, $D^{0}/\overline{D}^{0}$ and $%
B_{s}^{0}/\overline{B_{s}}^{0}$, neutral mesons\ and gives our conclusions.
In sections 2 and 4, $M^{0}/\overline{M}^{0}$ is either $K^{0}/%
\overline{K}^{0}$ or $B^{0}/\overline{B}^{0}$. In sections 5, 6 and 7 the
experimental numerical values used to evaluate the expressions are taken
from the most recent reference: {\it PDG 2024} \cite{5}.

\section{Mass eigenstates without CPV}

A generic flavored neutral meson state $\left| M\left( \tau \right)
\right\rangle $ is a functions of the meson proper time $\tau $ and a linear
superposition, with amplitudes $\left( a,b\right) $, of the two flavor
eigenstates $\left| M^{0}\right\rangle $ and $\left| \overline{M}%
^{0}\right\rangle $ ($K^{0}/\overline{K}^{0}$ or $B^{0}/\overline{B}^{0}$).
These flavor eigenstates provide an orthonormal basis of the mesons Hilbert
space: $\left\langle M^{0}\right. \left| M^{0}\right\rangle =\left\langle 
\overline{M}^{0}\right. \left| \overline{M}^{0}\right\rangle =1$ and $%
\left\langle \overline{M}^{0}\right. \left| M^{0}\right\rangle =0$. Mesons
states $\left| M\left( \tau \right) \right\rangle $ are unstable and decay
to a set of final states $\left| f\right\rangle $. Because of these decays
the Hilbert space should be extended to $\left\{ \left| f\right\rangle
\right\} $ and the time evolution should be described with an additional set
of amplitudes $\left\{ w_{f}\right\}$ as

\begin{equation}
\left| M\left( \tau \right) \right\rangle =a\left( \tau \right) \left|
M^{0}\right\rangle +b\left( \tau \right) \left| \overline{M}%
^{0}\right\rangle +\sum_{f}w_{f}\left( \tau \right) \left| f\right\rangle .
\label{a1}
\end{equation}

Following Lee, Oehme and Yang, the Weisskopf-Wigner (WW) approximation \cite
{13} is used here to describe the coupling to $\left| f\right\rangle $ as an
irreversible decay. Within the framework of this usual approximation \cite
{4,5}, rather than a unitary evolution with amplitudes $\left\{ w_{f}\left(
\tau \right) \right\} $, we introduce a non-Hermitian decay operator $j%
\widehat{\gamma }$ describing the ultimate $M\rightarrow f$ transitions as
an irreversible process. It is to be noted that $f\rightarrow M$ 
transitions, avoided by the WW approximation, are in fact experimentally 
prohibited due to the
very large phase space explored by the outgoing products $\left\{ \left| f\right\rangle
\right\} $.
The irreversibility of the decay process
gives rise to a violation of T that should not be attributed to fundamental interactions
at the level of the CKM matrix of the SM.

The time evolution of $\left|
M\left( \tau \right) \right\rangle $ can therefore be restricted to a two states
Hilbert space: $\left| M^{0}\right\rangle ,\left| \overline{M}%
^{0}\right\rangle $, at the cost of the loss of unitarity $d\left\langle
M\right. \left| M\right\rangle /d\tau <0$ induced by $j%
\widehat{\gamma }$. This restriction of the Hilbert space to ($
M^{0},\overline{M}^{0}$) leads to the
LOY Hamiltonian  $\widehat{H}_{Y}$ \cite{10,11}: the sum of the
mass energy ($mc^{2}$), plus a  $S=\pm 1$, or $
B^{\prime }=\pm 1$, mixing operator ($\widehat{\delta m}c^{2}$), plus the
irreversible decay:

\begin{equation}
\widehat{H}_{Y}=mc^{2}\widehat{I}-\widehat{\frac{\delta m}{2}}c^{2}-j\hbar 
\frac{\widehat{\gamma }}{2}\text{,}  \label{a2}
\end{equation}
where $\widehat{I}$ is the identity operator. The mixing and the decay
operators, $\widehat{\delta m}$ and $\widehat{\gamma }$, are given by 
\begin{eqnarray}
\widehat{\delta m} &=&\delta m\left[ \left| M^{0}\right\rangle \left\langle 
\overline{M}^{0}\right| +\left| \overline{M}^{0}\right\rangle \left\langle
M^{0}\right| \right] \text{,}  \label{a3} \\
\widehat{\gamma } &=&\Gamma \widehat{I}-\delta \Gamma \left[ \left|
M^{0}\right\rangle \left\langle \overline{M}^{0}\right| +\left| \overline{M}%
^{0}\right\rangle \left\langle M^{0}\right| \right] \text{,}  \label{a4}
\end{eqnarray}
where $\delta m>0$ is the mass splitting between the heavy and light mass
eigenstates and $\Gamma >0$, $\delta \Gamma <0$ are respectively the average
and the splitting between the decay widths of the these eigenstates \cite{4}%
. We take the convention $\widehat{CP}\left| M^{0}\right\rangle =\left| 
\overline{M}^{0}\right\rangle $. The evolution of the meson state $\left|
M\left( \tau \right) \right\rangle $ is  
\begin{equation}
j\hbar \frac{d\left| M\left( \tau \right) \right\rangle }{d\tau }=\widehat{H}%
_{Y}\cdot \left| M\left( \tau \right) \right\rangle \text{.}  \label{a125}
\end{equation}
The CP eigenstates $\left| M_{1}\right\rangle $ and $\left|
M_{2}\right\rangle $ are related to the flavor eigenstates by 
\begin{equation}
\left| M_{1/2}\right\rangle =\frac{\left| M^{0}\right\rangle }{\sqrt{2}}\pm 
\frac{\left| \overline{M}^{0}\right\rangle }{\sqrt{2}}=\pm \widehat{CP}%
\left| M_{1/2}\right\rangle \text{.}  \label{a6}
\end{equation}
These CP eigenstates are also mass eigenstates with eigenvalues $m\pm
\delta m/2$. 

The time evolution of these CP/mass eigenstates is given by the
solutions of (\ref{a125}): 
\begin{equation}
\left| M_{1/2}\left( \tau \right) \right\rangle =\left| M_{1/2}\right\rangle
\exp -j\frac{c^{2}}{\hbar }\left[ m\mp \frac{\delta m}{2}-j\hbar \frac{%
\Gamma \mp \delta \Gamma }{2c^{2}}\right] \tau \text{.}  \label{a8}
\end{equation}
The above symmetric picture, where $\widehat{CP}$ commute with $\widehat{H}%
_{Y}$, is no longer valid when the results of experiments dedicated to CPV are to be
taken into account. The observed mass eigenstates  are
not the CP  eigenstates $K_{1/2}$ or $B_{1/2}$\ (\ref{a6}). The mass
eigenvalues involved in (\ref{a8}) are not significantly changed by CPV. 
\section{Observable violation parameters}
The
observed mass eigenstates are: the short-lived $S$ and the long-lived $L$ 
states $\left( K_{S/L}\right) $ for $K^{0}/\overline{K}^{0}$, and the light $L$  
and heavy $H$ states $\left( B_{L/H}\right) $ for $B^{0}/\overline{B}^{0}$. 

For $K^{0}/\overline{K}^{0}$ type ({\it i}) CPV, the observed mass
eigenstates $\left| K_{S/L}\right\rangle $ are related to the CP eigenstates 
$\left| K_{1/2}\right\rangle $ (\ref{a6}) by 
\begin{equation} 
\left| K_{S/L}\right\rangle =\left| K_{1/2}\right\rangle +\varepsilon \left|
K_{2/1}\right\rangle \text{.}  \label{a10}
\end{equation}
The
quantity $\left\langle K_{S/L}\right. \left| K_{L/S}\right\rangle /2=%
\mathop{\rm Re}\varepsilon $ is an observable.

For $B^{0}/\overline{B}^{0}$ type ({\it iii}) CPV, it is convenient to
introduce the angle $\beta $ and to consider that the mass eigenstates $%
\left| B_{L/H}\right\rangle $ are related to the CP eigenstates $\left|
B_{1/2}\right\rangle $ (\ref{a6}) by 
\begin{equation}
\left| B_{L/H}\right\rangle =\cos \beta \left| B_{1/2}\right\rangle +j\sin
\beta \left| B_{2/1}\right\rangle \text{.}  \label{a12}
\end{equation}

Type ({\it ii}) direct CPV in the decay to one final state $\left\langle
f\right| $ is also due to Earth's gravity but $\varepsilon
^{\prime }$, the associated CPV parameter, is not involved in the LOY Hamiltonian describing $%
\Delta S=2$ oscillations. The measurements of the
direct violation parameter $\varepsilon ^{\prime }$ are based on difficult 
and precise dedicated pions
decays experiments. For the $2\pi ^{0}$ decays of $K_{L}$ and $K_{S}$ the
definition of $\varepsilon ^{\prime }$ is related to the amplitude ratio $%
\eta $ $_{00}$ by 
\begin{equation}
\eta _{00}=\frac{\left\langle \pi ^{0}\pi ^{0}\right| {\cal T}\left|
K_{L}\right\rangle }{\left\langle \pi ^{0}\pi ^{0}\right| {\cal T}\left|
K_{S}\right\rangle }\equiv \varepsilon -2\varepsilon ^{\prime }\text{,}
\end{equation}
where $\mathop{\rm Re}\varepsilon \gg \mathop{\rm Re}\varepsilon ^{\prime }$%
. CP symmetry is restored when $\varepsilon =0$, $\varepsilon ^{\prime }=0$
and $\beta =0$.

These experimental parameters, $\varepsilon $, $\varepsilon ^{\prime }$ and $%
\beta $,  are used to construct the CKM matrix elements. Rather than
adjusting the CPV part of the CKM matrix  to these measured parameters, a new interpretation of
the CPV experiments is proposed below. The final quantitative results
predicted with this new interpretation leads to the conclusion that CPV effects
observed in the three canonical types of flavored neutral mesons experiments
({\it i, ii} and {\it iii}) are gravity induced.

\section{Impact of Earth's gravity on neutral mesons oscillations}

We assume that the Schr\H{o}dinger equation  (\ref{a125}) is pertinent far from any
massive object, and, on Earth, we consider an additional energy term 
$mg\ \widehat{x}\left( \tau \right) $ in equation (\ref{a2}) such 
that the evolution becomes 
\begin{equation}
j\hbar \frac{d\left| M\right\rangle }{d\tau }=\widehat{H}_{Y}\cdot \left|
M\right\rangle +mg\ \ \widehat{x}\left( \tau \right) \cdot \left|
M\right\rangle \text{.}  \label{a13}
\end{equation}
The vertical position operator $\ \widehat{x}\left( \tau \right) $ is associated with
the vertical zitterbewegung internal motion inherent to all, free and bound,
spin 1/2 fermions like quarks \cite{9}.
At the level of the quark structure, the flavor eigenstates, 
$ M^{0} $ and $\overline{M}^{0} $, are stationary diquarks bound states 
($K^{0}/\overline{K}^{0}\sim \left( d\overline{s}\right) /\left( 
\overline{d}s\right) $ and $B^{0}/\overline{B}%
^{0}\sim \left( d\overline{b}\right) /\left( \overline{d}b\right)) $,
ultimately described by Dirac spinors, $
\left| q^{\prime }\overline{q}\right\rangle$ and $\left| \overline{q}^{\prime
}q\right\rangle $, for one light quark $q^{\prime }$ 
and one heavier quark $q$ combined into {\it singlet} spin zero states
: $M^{0} \sim \left| q^{\prime }\overline{q}\right\rangle $ and $%
\overline{M}^{0} $ $\sim $ $\left| \overline{q}^{\prime
}q\right\rangle $. It is to be noted that $\left|M^{0}\right\rangle/\left| 
\overline{M}^{0}\right\rangle $ and 
$\left| q^{\prime }\overline{q}\right\rangle/\left| \overline{q}^{\prime
}q\right\rangle $ belong to different Hilbert spaces.

A Dirac Hamiltonian $\widehat{H}_{D}$, describing quarks
confinement, operate in the diquark Hilbert space.
The position operator $%
\widehat{{\bf x}}\left( \tau \right) $, operating in the diquark spinor
Hilbert space, fulfils Heisenberg's equation: 
\begin{equation}
j\hbar \frac{d\widehat{{\bf x}}}{d\tau }=\left[ \widehat{{\bf x}},\widehat{H}%
_{D}\left( \widehat{{\bf x}},\widehat{{\bf p}}\right) \right] =j\hbar c{%
\boldsymbol \alpha }\text{.}  \label{a15}
\end{equation}
We have introduced the usual $4\times 4$ alpha matrices: ${\boldsymbol %
\alpha }=\left( \alpha _{x},\alpha _{y},\alpha _{z}\right) $  which
can be expressed in terms of the $2\times 2$ Pauli matrices ${\boldsymbol %
\sigma }=\left( \sigma _{x},\sigma _{y},\sigma _{z}\right) $. 

The non-zero matrix elements
of  ${\boldsymbol \alpha }c=d\widehat{{\bf x}}/d\tau$ are either $\pm c$ or $\pm jc$.  

It is important to note that the {\it zitterbewegung \ relation} $d%
\widehat{{\bf x}}/d\tau =c{\boldsymbol \alpha }$ (\ref{a15}) is independent of the
charge and mass of the fermions as well as of the shape and strength of the
effective confinement potential involved in the Dirac Hamiltonian $\widehat{H%
}_{D}$ \cite{9}.

A {\it zitterbewegung position operator} $\widehat{{\bf x}}\left( \tau
\right) $ operates also in the meson Hilbert space and is represented in
this space by the following (unknown) four matrix elements $\left\langle
{}\right| \widehat{{\bf x}}\left| {}\right\rangle $ of the stationary Dirac
spinors diquarks states $\left| q^{\prime }\overline{q}\right\rangle $ 
\begin{eqnarray}
\widehat{x }\left( \tau \right) &=&\left\langle q^{\prime }\overline{q}%
\right| \widehat{x}\left( \tau \right) \left| q^{\prime }\overline{q}%
\right\rangle \left| M^{0}\right\rangle \left\langle M^{0}\right|  \nonumber
\\
&&+\left\langle \overline{q}^{\prime }q\right| \widehat{x}\left( \tau
\right) \left| \overline{q}^{\prime }q\right\rangle \left| \overline{M}%
^{0}\right\rangle \left\langle \overline{M}^{0}\right|  \nonumber \\
&&+\left\langle \overline{q}^{\prime }q\right| \widehat{x}\left( \tau
\right) \left| q^{\prime }\overline{q}\right\rangle \left| \overline{M}%
^{0}\right\rangle \left\langle M^{0}\right|  \nonumber \\
&&+\left\langle q^{\prime }\overline{q}\right| \widehat{x}\left( \tau
\right) \left| \overline{q}^{\prime }q\right\rangle \left|
M^{0}\right\rangle \left\langle \overline{M}^{0}\right|
\text{.}  \label{a14}
\end{eqnarray}
The Compton wavelength of the meson $\lambda _{C}$ provides an approximate
upper bound of the matrix elements $\left| \left\langle {}\right| \widehat{x}%
\left| {}\right\rangle \right| $ in (\ref{a14}) as quarks are bound states
inside the volume of a meson. The small numerical value of the energy $%
mg\lambda _{C}=\hbar g/c\sim 10^{-23}$ eV in front of $\delta mc^{2}\sim
10^{-6}-10^{-4}$ eV leads to the occurrence of a strong ordering between $%
mg\left| \left\langle {}\right| \widehat{x}\left| {}\right\rangle \right| $ $%
\sim \hbar g/c$ $\ll \delta mc^{2}$ and the other LOY matrix elements
involved in $\widehat{H}_{Y}$. This ordering allows to set up a perturbative
expansion of \ (\ref{a13}) with respect to the small expansion parameter $%
\hbar g/\delta mc^{3}$ $\sim 10^{-19}-10^{-17}$. 

To do so we first define $%
\left| N\left( \tau \right) \right\rangle $ and $\left| n\left( \tau \right)
\right\rangle $ such that
\begin{equation}
\left| M\left( \tau \right) \right\rangle =\left| N\left( \tau \right)
\right\rangle \exp -j\frac{mc^{2}\tau }{\hbar }+\left| n\left( \tau \right)
\right\rangle \exp -j\frac{mc^{2}\tau }{\hbar }\text{.}
\end{equation}
The evolution of $\left| N\right\rangle +\left| n\right\rangle $ fulfils 
\begin{equation}
j\hbar \frac{d}{d\tau }\left[ \left| N \right\rangle
+\left| n \right\rangle \right] =\left[ \widehat{%
H^{\prime }}_{Y}+mg\ \ \widehat{x}\left( \tau \right) \right] \cdot \left[
\left| N \right\rangle +\left| n
\right\rangle \right] \text{,}\label{a144}
\end{equation}
where \ the Hamiltonian $\widehat{H^{\prime }}_{Y}$ is given by 
\begin{equation}
\widehat{H^{\prime }}_{Y}=\widehat{H}_{Y}-mc^{2}%
\widehat{I}=-\widehat{\delta m}c^{2}/2-j\hbar
\widehat{\gamma }/2\text{.} \end{equation}
Then the states $\left| N \right\rangle $ and $\left|
n\ \right\rangle $ are ordered according to: $\ \left| N%
\text{ }\right\rangle \sim $ $O\left( \hbar g/\delta mc^{3}\right) ^{0}$ and $%
O\left( \hbar g/\delta mc^{3}\right) ^{1}$ $\leq \left| n\right\rangle \ll
O\left( \hbar g/\delta mc^{3}\right) ^{0}$. With this expansion scheme Schr%
\"{o}dinger's equation (\ref{a144}) becomes 
\begin{eqnarray}
j\hbar \frac{d\left| N\right\rangle }{d\tau } &=&\widehat{H^{\prime }}%
_{Y}\cdot \left| N\right\rangle \text{,}  \label{a18} \\
j\hbar \frac{d\left| n\right\rangle }{d\tau } &=&\widehat{H^{\prime }}%
_{Y}\cdot \left| n\right\rangle +mg\ \widehat{x}\cdot \left| N\right\rangle 
\text{.}  \label{a19}
\end{eqnarray}
We introduce the inverse of the Hamiltonian $\widehat{H^{\prime }}_{Y}$ \ to
define the operators $\widehat{Z}_{x}$ and $\widehat{Z}_{c}=\hbar d\widehat{Z%
}_{x}/d\tau $%
\begin{eqnarray}
\widehat{Z}_{x}\left( \tau \right)  =jmg\ \widehat{x}\left(\tau \right)
\cdot \widehat{H^{\prime }}_{Y}^{-1},
\widehat{Z}_{c} =jmg\hbar \frac{d\widehat{{x}}}{d\tau } \cdot 
\widehat{H^{\prime }}_{Y}^{-1}\text{.}  \label{a457}
\end{eqnarray}
To evaluate the orders of magnitudes of $Z_{x}$ and $Z_{c}$ we ({\it i})
anticipate the specific cases of $K^{0}/\overline{K}^{0}$ and $B^{0}/%
\overline{B}^{0}$ where we will use respectively $\widehat{\gamma }_{K}=%
\widehat{0}$ and $\gamma _{B}\sim \delta m_{B}$ leading to $H_{Y}^{\prime
}\sim O\left( \delta mc^{2}\right) $ and ({\it ii)} use the fact that $d%
\widehat{{\bf x}}/d\tau =c{\boldsymbol \alpha }$ imply that the values of
the matrix elements of $d\widehat{x}/d\tau $ are independent of $\tau $ and
equal to $\pm c$, $\pm jc$ or $0$. 
The orders of magnitude of typical matrix
elements of $\widehat{Z}_{x}$ and $\widehat{Z}_{c}$, between two normalized
states, are thus $Z_{x}\sim O\left( \hbar g/\delta mc^{3}\right) $ and $%
Z_{c}\sim O\left( mc\hbar g/\delta mc^{2}\right) $.
With the definitions
(\ref{a457}) the relations (\ref{a18},\ref{a19}) give 
\begin{equation}
j\hbar \frac{d\left| n\right\rangle }{d\tau }=\widehat{H^{\prime }}_{Y}\cdot
\left| n\right\rangle -\widehat{Z}_{c}\ \cdot \left| N\right\rangle +\hbar 
\frac{d}{d\tau }\left[ \widehat{Z}_{x}\left( \tau \right) \cdot \left|
N\right\rangle \right] \text{.}  \label{a20}
\end{equation}
We introduce the state $\left| n^{\prime }\right\rangle =\left|
n\right\rangle +j\widehat{Z}_{x}\cdot \left| N\right\rangle $, such that 
\begin{equation}
j\hbar d\left| n^{\prime }\right\rangle /d\tau =\widehat{H^{\prime }}%
_{Y}\cdot \left| n^{\prime }\right\rangle -\left[ \widehat{Z}_{c}+j\widehat{%
H^{\prime }}_{Y}\cdot \widehat{Z}_{x}\right] \cdot \left| N\right\rangle 
\text{,}
\end{equation}
The operator $H_{Y}^{\prime }\cdot Z_{x}\sim O\left(\delta mc^{2}\right)\cdot O\left(
\hbar g/\delta mc^{3}\right) $ can be neglected in front $Z_{c}\sim
O\left( m\hbar g/\delta mc\right) $ as $\delta m/m\sim10^{-15}-10^{-13}$. 
To evaluate the interplay between 
zitterbewegung\ and $\Delta S=2$, or $\Delta B^{\prime }=2$, oscillations
we have to solve:
\begin{equation}
j\hbar \frac{d\left| n^{\prime }\right\rangle }{d\tau }=\widehat{H^{\prime }}%
_{Y}\cdot \left| n^{\prime }\right\rangle -\widehat{Z}_{c}\cdot \left|
N\right\rangle  \text{.}\label{a21}
\end{equation}
The last step is to express on the flavor basis $\left(
M^{0},\overline{M}^{0}\right) $ the
zitterbewegung instantaneous velocity operator 
$d\widehat{{x}}/d\tau$ involved in $%
\widehat{Z}_{c}$. As a consequence of (\ref{a15}) the eigenvectors of $d
\widehat{{\bf x}}/d\tau =c{\boldsymbol \alpha }$ can be identified. 
Without loss of
generality we consider the Dirac spinors eigenvectors of $\alpha
_{x}$: 
\begin{equation}
\frac{1}{\sqrt{2}}\left[ 
\begin{tabular}{c}
$1$ \\ 
$0$ \\ 
$0$ \\ 
$1$%
\end{tabular}
\right] \text{, }\frac{1}{\sqrt{2}}\left[ 
\begin{tabular}{c}
$0$ \\ 
$1$ \\ 
$1$ \\ 
$0$%
\end{tabular}
\right] \text{, }\frac{1}{\sqrt{2}}\left[ 
\begin{tabular}{c}
$1$ \\ 
$0$ \\ 
$0$ \\ 
$-1$%
\end{tabular}
\right] \text{, }\frac{1}{\sqrt{2}}\left[ 
\begin{tabular}{c}
$0$ \\ 
$1$ \\ 
$-1$ \\ 
$0$%
\end{tabular}
\right] \text{.}  \label{a123}
\end{equation}
The usual physical interpretation of these four spinors (\ref{a123}) is as
follows \cite{9}. 

Starting from the left, the first spinor and the second
one describe a symmetric superpositions of one fermion and one antifermion: $%
\left| g\right\rangle=\left( \left| q\right\rangle +\left| \overline{q}\right\rangle \right) /%
\sqrt{2}$. These two symmetric superpositions (\ref{a123}) are eigenstates
of $\alpha _{x}$ with eigenvalue $1$, and of $d\widehat{x}/d\tau$ with eigenvalue $+c$. 

The last two spinors, on the right,
describe an antisymmetric superpositions of one fermion and one antifermion: $%
\left| u\right\rangle=\left( \left| q\right\rangle -\left| \overline{q}\right\rangle \right) /%
\sqrt{2}$. These two antisymmetric superpositions (\ref{a123}) are
eigenstates of $\alpha _{x}$ with eigenvalue $-1$, and of $d\widehat{x}/d\tau$ with eigenvalue 
$-c$. 

Note that $\left\langle u\right. \left| g\right\rangle =0$. The symmetric 
CP eigenstate $ M_{1}$ (\ref{a6}) is
a combination of quarks spinors (\ref{a123}) of the $\left| g\right\rangle$ type and $
M_{2} $, the antisymmetric CP eigenstate (\ref{a6}), is
a combination of quarks spinors of the $\left| u\right\rangle$ type.
Thus, in the two states LOY Hilbert space, on the ($M_{1},M_{2}$) CP basis
(\ref{a6}), the representation of the zitterbewegung velocity operator $d%
\widehat{x}/d\tau $ is given by 
\begin{equation}
\frac{d\widehat{x}}{d\tau }=c\left| M_{1}\right\rangle \left\langle
M_{1}\right| -c\left| M_{2}\right\rangle \left\langle M_{2}\right| \text{.}
\label{a1622}
\end{equation}
On the flavor basis ($M^{0},\overline{M}^{0}$) (\ref{a1622}) gives the relations $%
\left\langle \overline{M}^{0}\right| d\widehat{x}/d\tau \left|
M^{0}\right\rangle $ = $c$ and $\left\langle M^{0}\right| d\widehat{x}/d\tau
\left| \overline{M}^{0}\right\rangle $ = $c$ and the two others matrix
elements are equal to zero. In the LOY Hamiltonian  (\ref{a2}) the mass $m
$ of the antiparticle is positive like the mass of the particle, although,
in a Dirac representation, the antiparticle are negative mass solutions.
This point is resolved through the Feynman interpretation of an antiparticle
as a particle propagating backward in time. To construct the LOY
representation of $d\widehat{x}/d\tau $, on the flavor basis ($M^{0},%
\overline{M}^{0}$), the usual Feynman prescription leads to the following
zitterbewegung velocity operator 
\begin{equation}
\frac{d\widehat{x}}{d\tau }=c\left| M^{0}\right\rangle \left\langle 
\overline{M}^{0}\right| -c\left| \overline{M}^{0}\right\rangle \left\langle
M^{0}\right| \text{.}  \label{a16}
\end{equation}
In two previous studies  \cite{12,14}, we have given two
demonstrations of this result (\ref{a16}) with two different
methods. Flavored neutral mesons pairs $K^{0}/\overline{K}^{0}$ and $B^{0}/%
\overline{B}^{0}$ display different $m$, $\delta m$, $\Gamma $ and $\delta
\Gamma $ and the impact of Earth gravity on their behavior is to be analyzed
specifically. In the following we keep the notations of (\ref{a457}, \ref{a21}) 
with an additional index $K$ or $B$ when needed.

\section{Type ({\it i}) CPV in the mixing of  $K^{0}/\overline{K}^{0}$}
The ordering associated with the specific case of a $K^{0}/\overline{K}^{0}$
pair is given by: $\delta m_{K}/m_{K}\sim 10^{-15}$. The first step to
interpret $K^{0}/\overline{K}^{0}$ experiments is to consider a unitary
evolution where both particles are regarded as stable, $j\hbar 
\widehat{\gamma }=\widehat{0}$. Then, as the lifetime of $K_{L}$
is 577 times longer than the lifetime of $K_{S}$, we will set up a
steady state-balance between: ({\it i}) the gravity induced small $
K_{S}$ component regenerated from a given $K_{L}$ and ({\it ii}) this $%
K_{S}$ component fast decay.

Considering first a unitary evolution, we have to solve (\ref{a18},\ref{a21}) 
\begin{equation}
j\hbar \frac{d\left| n^{\prime }\right\rangle }{d\tau }=-\frac{\widehat{%
\delta m_{K}}}{2}c^{2}\cdot \left| n^{\prime }\right\rangle
+2jm_{K}c^{-2}g\hbar \frac{d\widehat{x}}{d\tau }\ \cdot \widehat{\delta m_{K}}%
^{-1}\cdot \left| N\right\rangle   \label{a25}
\end{equation}
The operator $\widehat{\delta m_{K}}$ is given by (\ref{a3}) and the
operator $d\widehat{x}/d\tau $ by (\ref{a16}). 
As $\widehat{\delta m_{K}}\cdot \widehat{\delta m_{K}} =
\delta m_{K}^{2}\widehat{I}$
the actions of $\widehat{Z}_{c}$ and $\widehat{\delta m_{K}}$
on the CP eigenstates $\left| K_{1}\right\rangle $ and $\left|
K_{2}\right\rangle $ are 
\begin{eqnarray}
m_{K}g\hbar c^{-2}\frac{d\widehat{x}}{d\tau }\ \cdot \widehat{\delta m_{K}}%
^{-1}\cdot \left| K_{2/1}\right\rangle  &=&\kappa \frac{\delta m_{K}}{2}c^{2}%
\left| K_{1/2}\right\rangle \text{,}\nonumber
\\
\frac{\widehat{\delta m_{K}}}{2}\cdot \left| K_{2/1}\right\rangle  &=&\mp 
\frac{\delta m_{K}}{2}\left| K_{2/1}\right\rangle \text{,}
\end{eqnarray}
where the small parameter $\kappa $ is defined by 
\begin{equation}
\kappa =2m_{K}g\hbar /\delta m_{K}^{2}c^{3}=1.7\times 10^{-3}\text{.}
\label{a28}
\end{equation}
If we consider the following CP eigenstates, which are also the $\left(
m_{K}\pm \delta m_{K}/2\right) $ mass eigenstates without CPV 
\begin{equation}
\left| N\left( \tau \right) \right\rangle =\left| K_{2/1}\right\rangle \exp
\left( \mp j\delta m_{K}c^{2}\tau /2\hbar \right) \text{,}  \label{a29}
\end{equation}
they fulfils Eq. (\ref{a18}) and the associated solution of (\ref{a25})
is 
\begin{equation}
\left| n^{\prime }\left( \tau \right) \right\rangle =\pm j\kappa \left|
K_{1/2}\right\rangle \exp \left( \mp j\delta m_{K}c^{2}\tau /2\hbar \right) 
\text{.}  \label{a30}
\end{equation}
Thus, on Earth, the mass eigenstates $\left| K_{L/S}^{\oplus }\right\rangle $
are not the CP eigenstates $\left| K_{2/1}\right\rangle $, but  
\begin{equation}
\left| K_{L/S}^{\oplus }\right\rangle =\left| K_{2/1}\right\rangle \pm
j\kappa \left| K_{1/2}\right\rangle \text{.}  \label{a31}
\end{equation}
We neglect the $O[10^{-6}]$ correction needed for normalization 
$\left\langle K_{L/S}^{\oplus }\right. \left| K_{L/S}^{\oplus
}\right\rangle $ = $1$, and we have neglected the term $-j\widehat{Z}%
_{x}\cdot \left| K_{2/1}\right\rangle \sim O\left( \hbar g/\delta
m_{K}c^{3}\right) $ in front of $\kappa \left| K_{2/1}\right\rangle \sim
O\left( m_{K}\hbar g/\delta m_{K}^{2}c^{3}\right) $ as $\delta
m_{K}/m_{K}\sim 10^{-15}$. 
At the fundamental level of a unitary evolution,
without decays, the impact of Earth's gravity appears as a CPT violation,
with T conservation, \ because the indirect violation parameter $%
\left\langle K_{S}^{\oplus }\right. \left| K_{L}^{\oplus }\right\rangle $ = $%
2j\kappa $ is imaginary \cite{4}, rather than a CP and T violation with CPT
conservation requiring a non zero real value of $\left\langle K_{S}^{\oplus
}\right. \left| K_{L}^{\oplus }\right\rangle $ \cite{4}.

Usually, the three types of CPV experimental evidences are interpreted under
the assumption of CPT conservation. The CPT theorem is demonstrated within
the framework of three hypothesis: Lorentz group invariance, spin-statistics
relations and local field theory. In the rest frame of a meson interacting
with a massive spherical object, like Earth, the first hypothesis is not
satisfied. Thus, when Earth influence is taken into account, we must not be
surprised that CPT theorem, apparently, no longer holds. Within the
framework of a GICPV mechanism Earth's gravity is described as an external
field and the evolution of a meson state $\left| M\right\rangle $ alone, as
a linear superposition of two flavor eigenstates $\left| M^{0}\right\rangle $
and $\left| \overline{M}^{0}\right\rangle $, does not provide the complete
picture of the dynamical system and so can not be considered as a good
candidate displaying CPT invariance. Of course there are no CPT violation 
{\it stricto sensu,} CPT is restored for the global three bodies $\left(
M^{0}/\overline{M}^{0}/\oplus \right) $ quantum evolution of the state $%
\left| M\left( \tau \right) ,\oplus \right\rangle $ describing both the
meson-antimeson pair and Earth. 

In this study we consider only the evolution
of $\left| M\right\rangle $ and Earth's effect is described as an external
static field so that CPT will appear to be violated because of this
restricted two bodies $\left( M^{0}/\overline{M}^{0}\right) $ model of 
a three bodies system $%
\left( M^{0}/\overline{M}^{0}/\oplus \right) $. 

Moreover, with
GICPV there is no T violation at the microscopic level. 
As demonstrated in the next section, 
the
observed T violation stems from the irreversible decay of the short-lived
kaons $K_{S}$ continuously regenerated from the long-lived one $K_{L}$ by
gravity.

We must now take into account the $K_{S}$ fast decay. This decay will
change the picture, qualitatively: an apparent CP and T violation, with CPT
conservation rather than a CPT violation, and quantitatively: with the right
prediction of $\mathop{\rm Re}\varepsilon $.

The lifetime of the $K_{L}$ is 577 times larger than the lifetime of $\ K_{S}
$. The previous results (\ref{a31}) allows to calculate the{\it \
gravity induced transition rate} $\Omega _{L\rightarrow S}^{\oplus
}$ describing
the transition amplitude per unit time from the state $\left| K_{L}^{\oplus
}\right\rangle $ $\exp -j\delta m_{K}c^{2}\tau /2\hbar $ \ to the state $%
\left| K_{S}^{\oplus }\right\rangle $ 
$\exp j\delta m_{K}c^{2}\tau /2\hbar $:
\begin{equation}
\Omega _{L\rightarrow S}^{\oplus
}=\left\langle \frac{dK_{L}^{\oplus }}{d\tau }\left|
K_{S}^{\oplus }\left( \tau \right) \right. \right\rangle =\kappa \frac{%
\delta m_{K}c^{2}}{\hbar }\exp j\frac{\delta m_{K}c^{2}}{\hbar }\tau \text{.}
\label{a35}
\end{equation}
This can be viewed as a  {\it gravity induced regeneration}
competing with the short-lived kaon irreversible decay. This decay, to the
set of final states $\left\{ \left| f\right\rangle \right\} $, takes place
at a rate $\Gamma _{1\rightarrow f}/2=\left( \Gamma _{K}-\delta \Gamma
_{K}\right) /2$ $\sim $ $\Sigma _{f}\left| \left\langle f\right| {\cal T}%
\left| K_{1}\right\rangle \right| ^{2}$.

We consider now a typical experiment dedicated to indirect CPV.
Experimentally $K_{1}$ and $K_{2}$ are initially produced together in equal
amounts. Then, after few $1/$ $\Gamma _{1\rightarrow f}$ \ decay times, the
initial content of $\left| K_{1}\right\rangle $ disappears and a pure $%
\left| K_{2}\right\rangle $ state is expected. In fact, the state $\left|
K_{L\text{obs}}\left( \tau \right) \right\rangle $ observed in such an\
experiment is not a pure $\left| K_{2}\right\rangle $ state. This observed
state $\left| K_{L\ \text{obs}}\left( \tau \right) \right\rangle $ is a
linear superposition of $\left| K_{2}\right\rangle $, plus a small amount of 
$\left| K_{1}\right\rangle $, 
\begin{equation}
\left| K_{L\text{obs}}\left( \tau \right) \right\rangle =a_{2}\left( \tau
\right) \left| K_{2}\right\rangle +a_{1}\left( \tau \right) \left|
K_{1}\right\rangle \text{,}  \label{a36}
\end{equation}
resulting from the balance between gravity induced regeneration (\ref{a35})
and the fast irreversible decay of the $K_{1}$ component. We assume that the $K_{2}$
component is stable and that the depletion of its amplitude associated with
the {\it gravitational regeneration} of $K_{1}$ is negligible so that $%
\left| a_{2}\left( \tau \right) \right| $ = $1$ and 
\begin{equation}
a_{2}\left( \tau \right) =\exp -j\delta m_{K}c^{2}\tau /2\hbar \text{.}
\label{a37}
\end{equation}
The amplitude $a_{1}$ of $K_{1}$ in (\ref{a36}) is given by the steady-state
balance between a decay at the rate $\Gamma _{1\rightarrow f}/2$
on the one hand, and a gravity induced regeneration at the rate $\Omega _{L\rightarrow S}^{\oplus
}$ (\ref{a35}) from $K_{2}$ on the other hand. This steady-state balance reads 
\begin{equation}
a_{2}\left( \tau \right) \Omega _{L\rightarrow S}^{\oplus
}=a_{1}\left(\tau \right) 
\frac{\Gamma _{1\rightarrow f}}{2}\text{.}  \label{a38}
\end{equation}
The solution is this equation is 
\begin{equation}
a_{1}\left( \tau \right) =\frac{\delta m_{K}c^{2}}{\hbar \Gamma _{S}/2}%
\kappa \exp j\delta m_{K}c^{2}\tau /2\hbar \text{,}  \label{a39}
\end{equation}
where we have dropped the index $1\rightarrow f$ in $\Gamma$ to simplify the
notation and used $\Gamma _{S}$. The short-lived $\left| K_{1}\right\rangle $ component is
observed through its two pions decay \cite{1}. 

Thus the observed long-lived
mass eigenstate $\left| K_{L\text{obs}}\right\rangle $, obtained after few $%
1/$ $\Gamma _{S}$\ decay times away from a neutral kaons source, must be
represented by 
\begin{equation}
\left| K_{L\text{obs}}\right\rangle =\left| K_{2}\right\rangle +\frac{\delta
m_{K}c^{2}}{\hbar \Gamma _{S}/2}\kappa \left| K_{1}\right\rangle \text{.}
\label{a40}
\end{equation}
This is the usual CPV parametrization (\ref{a10}) of the mass eigenstates used under a CPT 
assumption.

The
observed value of the indirect GICPV parameter, 
\begin{equation}
\mathop{\rm Re}\varepsilon _{\text{obs}}=\frac{\delta m_{K}c^{2}}{\hbar
\Gamma _{S}/2}\frac{2m_{K}g\hbar }{\delta m_{K}^{2}c^{3}}=1.66\times 10^{-3}%
\text{,}  \label{a41}
\end{equation}
is in agreement with the most recent experimental value \cite{5}: 
\begin{equation}
\mathop{\rm Re}\varepsilon _{PDG2024}=\left( 1.66\pm 0.02\right) \times
10^{-3}\text{.} \label{a4111}
\end{equation}

To complete the previous analysis, we can also take into account the decay of the
other mass eigenstate, and this will reveal a {\it phenomenological dissipative
phase} of $\varepsilon $. 

Considering the decay rates $\Gamma _{S}=\Gamma
_{K}-\delta \Gamma _{K}$ for $K_{S}$, and $\Gamma _{L}$ = $\Gamma
_{K}+\delta \Gamma _{K}$ \ for $K_{L}$ $(\delta \Gamma _{K}<0$), beside their
usual definitions   in
terms of transition amplitudes, $\Gamma _{S/L}$ = $\Sigma _{f}\left|
\left\langle f\right| {\cal T}\left| K_{S/L}\right\rangle \right| ^{2}$, Bell and Steinberger  have
demonstrated a general relation based on global unitarity starting from the
evaluation of $\ d\left\langle M\right. \left| M\right\rangle /d\tau $ at $%
\tau =0$ \cite{15}. 

Using the fact that, for $K_{S}$, the sum $\Sigma _{f}$ \ over the
final states is dominated (99.9\%) by $K_{S}\rightarrow 2\pi $ decays, more
precisely by the $K_{S}$ $\rightarrow I_{0}$ decays (95\%) to the
isospin-zero combination of $\left| \pi ^{+}\pi ^{-}\right\rangle $ and $%
\left| \pi ^{0}\pi ^{0}\right\rangle $, the Bell-Steinberger's unitarity
relations  can be written \cite{15}: 
\begin{equation}
j\frac{\delta m_{K}c^{2}}{\hbar }+\frac{\Gamma _{S}}{2}=\frac{\left\langle
I_{0}\right| {\cal T}\left| K_{L}\right\rangle \left\langle I_{0}\right| 
{\cal T}\left| K_{S}\right\rangle ^{*}}{\left\langle K_{S}\right. \left|
K_{L}\right\rangle }\text{.}  \label{a42}
\end{equation}
The restriction of $\sum_{f}\left| f\right\rangle $ to $\left|
I_{0}\right\rangle $ reduces the $K_{S}$ width to $\Gamma _{S}$ = $%
\left\langle I_{0}\right| {\cal T}\left| K_{S}\right\rangle \left\langle
I_{0}\right| {\cal T}\left| K_{S}\right\rangle ^{*}$ so that 
\begin{equation}
\frac{\left\langle I_{0}\right| {\cal T}\left| K_{L}\right\rangle }{%
\left\langle I_{0}\right| {\cal T}\left| K_{S}\right\rangle }=\frac{%
\left\langle I_{0}\right| {\cal T}\left| K_{L}\right\rangle \left\langle
I_{0}\right| {\cal T}\left| K_{S}\right\rangle ^{*}}{\Gamma _{S}}\text{.}
\label{a43}
\end{equation}
This expression is then substituted in Bell-Steinberger's relation (\ref{a42}%
) to get the final expression 
\begin{equation}
\frac{\left\langle I_{0}\right| {\cal T}\left| K_{L}\right\rangle }{%
\left\langle I_{0}\right| {\cal T}\left| K_{S}\right\rangle }=\frac{%
\left\langle K_{S}\right. \left| K_{L}\right\rangle }{2}\left( 1+j\frac{%
2\delta m_{K}c^{2}}{\hbar \Gamma _{S}}\right) \text{.}  \label{a44}
\end{equation}
The left hand side of (\ref{a44}) can be considered as the definition of a
complex indirect CPV parameter $\varepsilon $ and $\left\langle K_{S}\right. \left|
K_{L}\right\rangle /2=\mathop{\rm Re}\varepsilon $ (\ref{a41}), thus the argument of this
CPV complex parameter $\varepsilon $ is:  
\begin{equation}
\arg \varepsilon =\arctan \left( 2\delta m_{K}c^{2}/\hbar \Gamma _{S}\right)
=43.4^{\circ }\text{,}  \label{a45}
\end{equation}
in agreement with the experimental result $43.5^{\circ }$ \cite{5}. This
last relation (\ref{a45}) complements (\ref{a41}) and confirms that GICPV  
provides a global and pertinent framework to interpret $K^{0}/%
\overline{K}^{0}$ indirect CPV experiments.

It is very important to note that the fundamental parameter describing
indirect CPV is associated with the unitary evolution overlap of the mass
eigenstates induced by Earth's gravity: 
\begin{equation}
\frac{\left\langle K_{S}^{\oplus }\right. \left| K_{L}^{\oplus
}\right\rangle }{2}=j\frac{2m_{K}g\hbar }{\delta m_{K}^{2}c^{3}}\text{,}
\label{a4533}
\end{equation}
and, as explained above, the measurements of the complex CPV parameter given
by 
\begin{equation}
\varepsilon =\frac{2m_{K}g\hbar }{\delta m_{K}^{2}c^{3}}\left[ \frac{2\delta m_{K}c^{2}}{%
\hbar \Gamma _{S}}\left( 1+j\frac{2\delta m_{K}c^{2}}{\hbar \Gamma _{S}}%
\right) \right] \text{,}  \label{a4522}
\end{equation}
is due to a {\it dissipative dressing} of the fundamental overlap (\ref{a4533}),
{\it dissipative dressing} resulting from the finite lifetime of the mesons. This dissipative dressing
is not {\it stricto sensu} a CPV effects but is inherent to experiments with
unstable particles, this point is important to interpret CPV experiments 
and to understand the nature of GICPV.

\section{Type ({\it ii}) CPV in the decay of $K^{0}/\overline{K}^{0}$}

The analysis of type ({\it ii}) CPV in the decay to a final state $%
\left\langle f\right|$
rely on the measurement of the ratio $%
\eta _{f}=\left\langle f\right| {\cal T}\left| K_{L}\right\rangle
/\left\langle f\right| {\cal T}\left| K_{S}\right\rangle$.
To interpret the measurements of the direct violation parameter $\varepsilon
^{\prime }$ we consider $\left\langle f\right| =\left\langle \pi ^{0}\pi
^{0}\right| $ and the $2\pi ^{0}$ decays of $K_{L}$ and $K_{S}$ \cite{16,17,18}. The
definition of the direct CPV parameter $\varepsilon ^{\prime }$, as a
function of the amplitude ratio $\eta _{00}$, is given by 
\begin{equation}
\eta _{00}=\frac{\left\langle \pi ^{0}\pi ^{0}\right| {\cal T}\left|
K_{L}\right\rangle }{\left\langle \pi ^{0}\pi ^{0}\right| {\cal T}\left|
K_{S}\right\rangle }\equiv \varepsilon -2\varepsilon ^{\prime }\text{.}
\label{a46}
\end{equation}

The various bra and ket in a quantum model are defined up to an unobservable
phase. The arbitrary conventional phases inherent to quantum theoretical
models are to be eliminated to define {\it phase-convention-independent}
observables. 

The definition of $\eta _{00}$ is invariant under rephasing of
the pions state $\left\langle \pi ^{0}\pi ^{0}\right| $, but not with
respect to the rephasing of the kaons mass eigenstates $\left|
K_{L/S}\right\rangle $. We can define a decay amplitude ratio which is a
{\it phase-convention-independent} quantity through the multiplication of $\eta
_{00}$ with the {\it rephasing} factor $\varphi _{K}$ 
\begin{equation}
\varphi _{K}=\frac{\left\langle K^{0}\right. \left| K_{S}\right\rangle }{%
\left\langle K^{0}\right. \left| K_{L}\right\rangle }.  \label{a47}
\end{equation}
Within the SM-CKM framework, CPT invariance is assumed and the observed mass
eigenstates (\ref{a10}) are parametrized as: 
\begin{equation}
\left| K_{S/L}\right\rangle =\frac{1+\varepsilon }{\sqrt{2}}\left|
K^{0}\right\rangle \pm \frac{1-\varepsilon }{\sqrt{2}}\left| \overline{K}%
^{0}\right\rangle \text{.}  \label{a49}
\end{equation}
Within the GICPV framework, the mass eigenstates (\ref{a31})
display a different structure and are given by: 
\begin{equation}
\left| K_{S/L}^{\oplus }\right\rangle =\frac{1\mp j\kappa }{\sqrt{2}}\left|
K^{0}\right\rangle \pm \frac{1\pm j\kappa }{\sqrt{2}}\left| \overline{K}%
^{0}\right\rangle .  \label{a490}
\end{equation}
For the usual CPV parametrization (\ref{a49}) we obtain
\begin{equation}
\varphi _{K}=\frac{\left\langle K^{0}\right. \left| K_{S}\right\rangle }{%
\left\langle K^{0}\right. \left| K_{L}\right\rangle }=1\text{.}  \label{a52}
\end{equation}
For GICPV (\ref{a490}) we obtain 
\begin{equation}
\varphi _{K}^{\oplus }=\frac{\left\langle K^{0}\right. \left| K_{S}^{\oplus
}\right\rangle }{\left\langle K^{0}\right. \left| K_{L}^{\oplus
}\right\rangle }=1-\left\langle K_{S}^{\oplus }\right. \left| K_{L}^{\oplus
}\right\rangle \text{,}  \label{a 53}
\end{equation}
where $O\left[ 10^{-6}\right] $ and higher orders terms are neglected.

The interaction between a $\left( \pi ^{0},\pi ^{0}\right) $ state and a
neutral kaon state, $K^{0}$ or $\overline{K}^{0}$, can not differentiate the 
$K^{0}$ from the $\overline{K}^{0}$ (a final state phase can be absorbed by
a proper phase convention between $K^{0}$ and $\overline{K}^{0}$), thus the
amplitude of $K^{0}\rightarrow \pi ^{0}\pi ^{0}$ can be taken to be equal to
the amplitude of $\overline{K}^{0}\rightarrow \pi ^{0}\pi ^{0}$. 

Using equation (\ref{a490}),
the ratio of amplitudes $\eta _{00}^{\oplus }$ associated with
the unitary mass eigenstates resulting from GICPV is 
\begin{equation}
\eta _{00}^{\oplus }=\frac{\left\langle \pi ^{0}\pi ^{0}\right| {\cal T}%
\left| K_{L}^{\oplus }\right\rangle }{\left\langle \pi ^{0}\pi ^{0}\right| 
{\cal T}\left| K_{S}^{\oplus }\right\rangle }=\frac{\left\langle
K_{S}^{\oplus }\right. \left| K_{L}^{\oplus }\right\rangle }{2}.  \label{a54}
\end{equation}
We conclude that the (unitary $ 
\widehat{\gamma_{K} }=\widehat{0}$) physical observable $\eta _{00}^{\oplus }\varphi
_{K}^{\oplus }$ is given by 
\begin{equation}
\eta _{00}^{\oplus }\varphi _{K}^{\oplus }=\frac{\left\langle K_{S}^{\oplus
}\right. \left| K_{L}^{\oplus }\right\rangle }{2}\left[ 1-2\frac{%
\left\langle K_{S}^{\oplus }\right. \left| K_{L}^{\oplus }\right\rangle }{2}%
\right] \text{.}  \label{a55}
\end{equation}
This is the GICPV physical structure of the phase-convention-independent
amplitude ratio $\eta _{00}$.  However, finite lifetimes and decays are inherent to the 
experiments, this results in a measured amplitude ratio 
\begin{equation}
\eta _{00\text{obs}}^{\oplus }\varphi _{K\text{obs}}^{\oplus }=\frac{%
\left\langle K_{S}\right. \left| K_{L}\right\rangle }{2}\left[ 1-2\frac{%
\left\langle K_{S}\right. \left| K_{L}\right\rangle }{2}\right] .
\label{a56}
\end{equation}
Using the phase-convention-independent definition of $\varepsilon ^{\prime }$ (\ref{a46}) 
within the SM-CKM framework, the measurement 
of $\eta _{00}$ is normally interpreted as a direct CPV
\begin{equation}
\eta _{00\text{obs}}^{\oplus }\varphi _{K\text{obs}}^{\oplus }=\varepsilon
\left[ 1-2\frac{\varepsilon ^{\prime }}{\varepsilon }\right] \text{.}
\label{b7}
\end{equation}
The relations (\ref{a56}) and (\ref{b7}) lead to the conclusion $\mathop{\rm
Re}\left( \varepsilon ^{\prime }/\varepsilon \right) =\mathop{\rm Re}\left(
\varepsilon \right) $. 
The GI direct CPV parameter 
\begin{equation}
\mathop{\rm Re}\left( \varepsilon ^{\prime }/\varepsilon \right) _{\text{%
GICPV}}=\frac{\delta m_{K}c^{2}}{\hbar \Gamma _{S}/2}\kappa =1.66\times
10^{-3},  \label{a57}
\end{equation}
is in agreement with the most recent experimental value \cite{5}: 
\begin{equation}
\mathop{\rm Re}\left( \varepsilon ^{\prime }/\varepsilon \right) _{\text{%
{\it PDG2024}}}=\left( 1.66\pm 0.23\right) \times 10^{-3}.  \label{b8}
\end{equation}
The fact that $\mathop{\rm Re}\left( \varepsilon ^{\prime }/\varepsilon
\right) \sim \mathop{\rm Re}\left( \varepsilon \right) $ was considered, up
to now, as a numerical coincidence and it finds here a simple explanation in
term of a phase-convention-independent amplitude ratio within the framework
of GICPV.

The precise definition of phase-convention-independent quantities, in order
to clearly identify what is measured in an experiment, is also one of the
key to interpret the experimental observation of interferences between
mixing and decay in $B^{0}/\overline{B}^{0}$ CPV dedicated  experiments.

\section{Type ({\it iii}) CPV in the interference between mixing and decay}

Up to 2001, the evidences of CPV where restricted to K mesons experiments
and to the baryons asymmetry of the universe. In 2001 the first clear
identification of CPV with B mesons experiments in B-factories was reported 
\cite{19, 20}. The mass and width ordering associated with the $B^{0}/%
\overline{B}^{0}$ system are given by : $\delta m_{B}/m_{B}\sim 10^{-19}$
and $\delta m_{B}/\Gamma _{B}\sim 0.7$. The lifetime of the CP eigenstate $%
B_{1}$ is considered to be equal to the lifetime of the other CP eigenstate $%
B_{2}$ so that $\delta \Gamma _{B}=0$. 

The most pronounced CPV effects in
the $B^{0}/\overline{B}^{0}$ system are displayed through interference
experiments dedicated to the study of the phase difference between the decay
path $B_{0}\rightarrow f$ and the decay path $B_{0}\rightarrow \overline{B}%
^{0}\rightarrow f$ \ \cite{21, 22, 23}.

To set up an interpretation of these experiments we keep a finite lifetime $%
\Gamma _{B}^{-1}$ for both particles and consider the decay operator 
\begin{equation}
\widehat{\gamma }_{B}=\Gamma _{B}\left[ \left| B^{0}\right\rangle
\left\langle B^{0}\right| +\left| \overline{B}^{0}\right\rangle \left\langle 
\overline{B}^{0}\right| \right] \text{,}  \label{a58}
\end{equation}
to describe the dissipative part of \ the $B^{0}/\overline{B}^{0}$ dynamics.
Thus, we have to solve  (\ref{a21} ) 
\begin{equation}
j\hbar \frac{d\left| n^{\prime }\right\rangle }{d\tau }=\widehat{H^{\prime }}%
_{YB}\cdot \left| n^{\prime }\right\rangle -jm_{B}g\hbar \frac{d\widehat{x}}{%
d\tau }\ \cdot \widehat{H^{\prime }}_{YB}^{-1}\cdot \left| N\right\rangle 
\text{.}  \label{a60}
\end{equation}
The action of $\ jm_{B}g\hbar \frac{d\widehat{x}}{d\tau }\ \cdot \widehat{%
H^{\prime }}_{YB}^{-1}$ on the CP eigenstates $\left| B_{1}\right\rangle $
and $\left| B_{2}\right\rangle $ is 
\begin{equation}
\ jm_{B}g\hbar \frac{d\widehat{x}}{d\tau }\ \cdot \widehat{H^{\prime }}%
_{YB}^{-1}\left| B_{2/1}\right\rangle =\pm \delta m_{B}c^{2}\varsigma \left(
1\mp j\chi \right) \left| B_{1/2}\right\rangle \text{.}  \label{a62}
\end{equation}
where we have defined the real parameters $\chi=0.77 $ and 
$\varsigma\sim O\left[ 10^{-6}\right] $ as
\begin{eqnarray}
\chi  =\delta m_{B}c^{2}/\hbar \Gamma _{B}\text{, }  
\varsigma  =2m_{B}g\hbar /\delta m_{B}^{2}c^{3}\left( \chi +\chi
^{-1}\right)  \text{.}  \label{a64}
\end{eqnarray}
In order to solve equation (\ref{a60}) and to express the mass eigenstates on
Earth, we consider the CP eigenstates 
\begin{equation}
\left| N\left( \tau \right) \right\rangle =\left| B_{2/1}\right\rangle \exp
\mp j\frac{\delta m_{B}c^{2}\mp j\hbar \Gamma _{B}}{2\hbar }\tau \text{,}
\label{a65}
\end{equation}
which are also $\left( m_{B}\pm \delta m_{B}/2\right) $ mass eigenstate
without CPV.
The associated solution of \ (\ref{a60}) is 
\begin{equation}
\left| n'\left( \tau \right) \right\rangle =-\varsigma \left( 1\mp j\chi
\right) \left| B_{1/2}\right\rangle \exp \mp j\frac{\delta m_{B}c^{2}\mp
j\hbar \Gamma _{B}}{2\hbar }\tau \text{.}  \label{a66}
\end{equation}
Thus, on Earth, the mass eigenstates $\left| B_{L/H}^{\oplus }\right\rangle $
are not the CP eigenstates $\left| B_{1/2}\right\rangle $, but 
\begin{equation}
\left| B_{L/H}^{\oplus }\right\rangle =\left| B_{1/2}\right\rangle
-\varsigma \left( 1\pm j\chi \right) \left| B_{2/1}\right\rangle \text{.}
\label{a70}
\end{equation}
Using the flavor basis $\left[ \left| B^{0}\right\rangle ,\left| \overline{B}%
^{0}\right\rangle \right] $, rather than the CP basis $\left[ \left|
B_{1}\right\rangle ,\left| B_{2}\right\rangle \right] $, these mass
eigenstates (\ref{a70}) are expressed as 
\begin{equation}
\left| B_{L/H}^{\oplus }\right\rangle =\frac{1-\varsigma \left( 1\pm j\chi
\right) }{\sqrt{2}}\left| B^{0}\right\rangle \pm \frac{1+\varsigma \left(
1\pm j\chi \right) }{\sqrt{2}}\left| \overline{B}^{0}\right\rangle \text{.}
\label{a72}
\end{equation}

This GICPV results requires two real number, $\varsigma $ and $\chi $, to
express the mass eigenstates $\left| B_{L/H}^{\oplus }\right\rangle $ although
type ({\it iii}) standard SM-CKM parametrization (\ref{a12}) is based on a
single angle $\beta $ and display a different structure 
\begin{equation}
\left| B_{L/H}\right\rangle =\frac{\exp +j\beta }{\sqrt{2}}\left|
B^{0}\right\rangle \pm \frac{\exp -j\beta }{\sqrt{2}}\left| \overline{B}%
^{0}\right\rangle \text{.}  \label{a74}
\end{equation}

To interpret the CPV experiments we consider the decay into one final 
CP eigenstate $\left| f\right\rangle $.
The amplitude ratio $\lambda _{f}$ 
\begin{equation}
\lambda _{f}=\frac{\left\langle \overline{B}^{0}\right. \left|
B_{L}\right\rangle }{\left\langle B^{0}\right. \left| B_{L}\right\rangle }%
\frac{\left\langle f\right| {\cal T}\left| \overline{B}^{0}\right\rangle }{%
\left\langle f\right| {\cal T}\left| B^{0}\right\rangle }=\exp -2j\beta 
\frac{\left\langle f\right| {\cal T}\left| \overline{B}^{0}\right\rangle }{%
\left\langle f\right| {\cal T}\left| B^{0}\right\rangle }\text{,}
\label{a75}
\end{equation}
is observable within the SM-CKM framework because the SM-CKM 
parametrization (\ref{a74})
is reduced to a single angle which leads to the relations:
\begin{eqnarray}
\frac{\left\langle \overline{B}^{0}\right. \left| B_{L}\right\rangle }{%
\left\langle B^{0}\right. \left| B_{L}\right\rangle }=-\frac{\left\langle 
\overline{B}^{0}\right. \left| B_{H}\right\rangle }{\left\langle
B^{0}\right. \left| B_{H}\right\rangle }=\frac{\left\langle \overline{B}%
^{0}\right. \left| B_{L}\right\rangle }{\left\langle B^{0}\right. \left|
B_{H}\right\rangle } \nonumber \\  
=-\frac{\left\langle \overline{B}^{0}\right. \left|
B_{H}\right\rangle }{\left\langle B^{0}\right. \left| B_{L}\right\rangle }
=\exp -2j\beta%
\text{.}  \label{a77}
\end{eqnarray}
However, these four amplitudes ratios are different if we consider the
gravity induced mass eigenstates (\ref{a72}) 
\begin{equation}
\frac{\left\langle \overline{B}^{0}\right. \left| B_{L}^{\oplus
}\right\rangle }{\left\langle B^{0}\right. \left| B_{L}^{\oplus
}\right\rangle }\neq -\frac{\left\langle \overline{B}^{0}\right. \left|
B_{H}^{\oplus }\right\rangle }{\left\langle B^{0}\right. \left|
B_{H}^{\oplus }\right\rangle }\neq \frac{\left\langle \overline{B}%
^{0}\right. \left| B_{L}^{\oplus }\right\rangle }{\left\langle B^{0}\right.
\left| B_{H}^{\oplus }\right\rangle }\neq -\frac{\left\langle \overline{B}%
^{0}\right. \left| B_{H}^{\oplus }\right\rangle }{\left\langle B^{0}\right.
\left| B_{L}^{\oplus }\right\rangle }\text{.}  \label{a78}
\end{equation}

Despite this difference between (\ref{a77}) and (\ref{a78}), the
experimental results analyzed within a SM-CKM framework
can be understood and explained within the framework of GICPV. This situation is similar to the one encountered in the previous
section devoted to the study of $\varepsilon ^{\prime }$: if CPT is assumed
the rephasing factors $\varphi =1$, and the interpretation of the
experimental measurements is based on the hypothesis of direct violation and
imply a CPV at the fundamental level of the CKM matrix. However, if Earth's
gravity effects are taken into account $\varphi \neq 1$ and the very same
phase-convention-independent measured quantities agree with the experiments
without any additional assumption. 

The analysis below will use two different approaches to interpret the
measurement of $\ \beta $, each providing the same final result. 

The two issues addressed below are: ({\it i}) the invariance under rephasing
of the mass eigenstates to define an observable and ({\it ii})
the invariance under rephasing of the flavor eigenstates to
define an observable.

In order to accommodate the relation (\ref{a75}) with (\ref{a77}, \ref{a78}%
), we consider a $\widetilde{\lambda }
_{f}$ parameter constructed with the amplitude
ratio $\left\langle \overline{B}^{0}\right. \left| B_{L}\right\rangle
/\left\langle B^{0}\right. \left| B_{H}\right\rangle $ which is better
suited to characterize the dynamics of oscillating $B_{L/S}$ as it takes
into account all the eigenstates: the two flavor eigenstates and the two
mass eigenstates involved in experiments. However, this $\widetilde{\lambda }%
_{f}$ parameter: 
\begin{equation}
\widetilde{\lambda }_{f}=\frac{\left\langle \overline{B}^{0}\right. \left|
B_{L}\right\rangle }{\left\langle B^{0}\right. \left| B_{H}\right\rangle }%
\frac{\left\langle f\right| {\cal T}\left| \overline{B}^{0}\right\rangle }{%
\left\langle f\right| {\cal T}\left| B^{0}\right\rangle }=\lambda _{f}\text{,}  \label{a79}
\end{equation}
is not phase-convention-independent with respect to the mass eigenstates. 

To
set up a fully phase-convention-independent parameter we introduce the
symmetric rephasing factor $\varphi _{B}$: 
\begin{equation}
\varphi _{B}=\sqrt{\frac{\left\langle B_{1}\right. \left| B_{H}\right\rangle 
}{\left\langle B_{1}\right. \left| B_{L}\right\rangle }\frac{\left\langle
B_{2}\right. \left| B_{H}\right\rangle }{\left\langle B_{2}\right. \left|
B_{L}\right\rangle }}=1\text{.}  \label{a80}
\end{equation}
We have used $B_{1/2}$ states because they are CP eigenstates like $f$. The
amplitude ratio observed in the experimental measurement are given by
phase-convention-independent product $\widetilde{\lambda }_{f}\varphi _{B}$ 
\begin{equation}
\widetilde{\lambda }_{f}\varphi _{B}=\frac{\left\langle \overline{B}%
^{0}\right. \left| B_{L}\right\rangle }{\left\langle B^{0}\right. \left|
B_{H}\right\rangle }\frac{\left\langle f\right| {\cal T}\left| \overline{B}%
^{0}\right\rangle }{\left\langle f\right| {\cal T}\left| B^{0}\right\rangle }%
\varphi _{B}=\exp -2j\beta \frac{\left\langle f\right| {\cal T}\left| 
\overline{B}^{0}\right\rangle }{\left\langle f\right| {\cal T}\left|
B^{0}\right\rangle }  \label{a81}
\end{equation}
which is equal to $\lambda _{f}$ (\ref{a75}).

When the same rephasing factor 
$\varphi _{B}^{\oplus }$ is calculated within the framework of GICPV with (\ref{a70})  this gives 
\begin{equation}
\varphi _{B}^{\oplus }=\sqrt{\frac{\left\langle B_{1}\right. \left|
B_{H}^{\oplus }\right\rangle }{\left\langle B_{1}\right. \left|
B_{L}^{\oplus }\right\rangle }\frac{\left\langle B_{2}\right. \left|
B_{H}^{\oplus }\right\rangle }{\left\langle B_{2}\right. \left|
B_{L}^{\oplus }\right\rangle }}=\sqrt{\frac{1-j\chi }{1+j\chi }}\text{.}
\label{a82}
\end{equation}
The phase-convention-independent product 
$\widetilde{\lambda }_{f}^{\oplus }\varphi _{B}^{\oplus }$ is defined as
\begin{equation}
\widetilde{\lambda }_{f}^{\oplus }\varphi _{B}^{\oplus }=\frac{\left\langle 
\overline{B}^{0}\right. \left| B_{L}^{\oplus }\right\rangle }{\left\langle
B^{0}\right. \left| B_{H}^{\oplus }\right\rangle }\frac{\left\langle
f\right| {\cal T}\left| \overline{B}^{0}\right\rangle }{\left\langle
f\right| {\cal T}\left| B^{0}\right\rangle }\varphi _{B}^{\oplus }\text{,}
\end{equation}
and is given by
\begin{equation}
\widetilde{\lambda }_{f}^{\oplus }\varphi _{B}^{\oplus }=
\exp \left(
-j\arctan \chi \right) \frac{\left\langle f\right| {\cal T}\left| \overline{B%
}^{0}\right\rangle }{\left\langle f\right| {\cal T}\left| B^{0}\right\rangle 
}\left( 1+O\left[ 10^{-6}\right] \right) \text{.}  \label{a84}
\end{equation}

To compare the interpretations based on the usual SM-CKM eigenstates $\left|
B_{L/H}\right\rangle $ (\ref{a74}) with the gravity induced mass eigenstates 
$\left| B_{L/H}^{\oplus }\right\rangle $ (\ref{a72}), we must define $\beta $
such that $2\beta =$ $\arctan \left( 0.77\right) $. If $\left\langle
f\right| {\cal T}\left| \overline{B}^{0}\right\rangle /\left\langle f\right| 
{\cal T}\left| B^{0}\right\rangle $ is assumed real and equal to one the
experiments dedicated to the measure of $\lambda _{f}$ give a measurement equal to 
\begin{equation}
\sin 2\beta =\sin \left[ \arctan \left( 0.77\right) \right] =0.61\text{.}  
\label{a184}
\end{equation}
The modes $\overline{b}\rightarrow \overline{s}s\overline{s}$ and $\overline{%
b}\rightarrow \overline{c}c\overline{s}$ have been studied in depth through $B_{0}\rightarrow \phi K_{S}^{0}$ and $B_{0}\rightarrow \psi
K^{0}$ interferences. According to the data reported in \cite{5}
the present status of the values is 
$\sin 2\beta _{\phi K_{S}^{0}}=0.58\pm 0.12\text{,   }
\sin 2\beta _{\psi
K^{0}}=0.701\pm 0.01$. 
Other neutral final states, such as $J/\psi K^{*0}$ and $K^{0}\pi ^{0}$,
giving $0.60\pm 0.24\pm 0.08$ and $0.64\pm 0.13$, are in good agreement 
with the GICPV result (\ref{a184}) if $%
\left\langle f\right| {\cal T}\left| \overline{B}^{0}\right\rangle $ = $%
\left\langle f\right| {\cal T}\left| B^{0}\right\rangle $. For the full
set of final states $f$ studied up to now, the results are centered around (\ref
{a184}) but deviate from this value. The difficulty to evaluate $\arg
\left\langle f\right| {\cal T}\left| \overline{B}^{0}\right\rangle
/\left\langle f\right| {\cal T}\left| B^{0}\right\rangle $ is one source of
the dispersion, note also that the sign of $\left\langle
f\right| \widehat{CP}\left| f\right\rangle $ is to be considered  and the fact that (\ref{a74}) is assumed rather than (%
\ref{a72}) is probably also a source of dispersion. A clear understanding of
the $\sin 2\beta $ distribution around $0.6-0.7$ requires to drop (\ref{a74}%
) and to adopt the mass eigenstates (\ref{a72}); a precise evaluation of $\ \left\langle f\right| {\cal T}%
\left| \overline{B}^{0}\right\rangle /\left\langle f\right| {\cal T}\left|
B^{0}\right\rangle $ is also needed.

Let us now consider a second point of view. We will not consider the
interpretation of interferences experiments and, rather than addressing the
issue of $\lambda _{f}$, we address directly the issue of $\beta $ through a gedanken experiment. 
We consider the different mass eigenstates expansions on either CP or flavor
eigenstates: (\ref{a12}, \ref{a74}) for the SM-CKM framework, and (\ref{a70}, \ref
{a72}) for the GICPV framework. 

In order to compare the usual eigenstates
parametrization (\ref{a74}), based on a single angle $\beta $, with the
gravity induced mass eigenstates (\ref{a72}), involving two parameters $%
\varsigma $ and $\chi $, we must define $\beta $ through a {\it gedanken}
experiment providing $\exp j\beta $ as a phase-convention-independent
expression. We consider the symmetric and complete combination 
\begin{equation}
\rho _{B}=\frac{\left\langle B^{0}\right. \left| B_{L}\right\rangle }{%
\left\langle \overline{B}^{0}\right. \left| B_{L}\right\rangle }\frac{%
\left\langle B^{0}\right. \left| B_{H}\right\rangle }{\left\langle \overline{%
B}^{0}\right. \left| B_{H}\right\rangle }\text{,}  \label{a86}
\end{equation}
which takes into account the four components at work in the description.
This definition of $\beta $ through $\rho _{B}$ takes into account all
flavor and mass eigenstates but suffers from a lack of (unphysical) phase
compensation with respect to the flavor eigenstates. All measured
observables, independently of the interpretation of the measurement, are
combinations of phase-convention-independent quantities. We introduce the
coefficient $\varphi _{B}^{\prime }$ needed to provide a
phase-convention-independent observable associated with $\rho _{B}$%
\begin{equation}
\varphi _{B}^{\prime }=\frac{\left\langle \overline{B}^{0}\right. \left|
B_{2}\right\rangle \left\langle B_{2}\right. \left| B_{H}\right\rangle }{%
\left\langle B^{0}\right. \left| B_{1}\right\rangle \left\langle
B_{1}\right. \left| B_{L}\right\rangle }\frac{\left\langle \overline{B}%
^{0}\right. \left| B_{2}\right\rangle \left\langle B_{2}\right. \left|
B_{L}\right\rangle }{\left\langle B^{0}\right. \left| B_{1}\right\rangle
\left\langle B_{1}\right. \left| B_{H}\right\rangle }\text{,}  \label{a87}
\end{equation}
where we have chosen the two projection operators $\left| B_{1}\right\rangle
\left\langle B_{1}\right| $ and $\left| B_{2}\right\rangle \left\langle
B_{2}\right| $ because they commute with CP.

It can be checked that the product $\rho _{B}\varphi _{B}^{\prime }$ is
phase-convention-independent and thus can be measured in a gedanken
experiment which does not need to be described here.

If the usual parametrization of CPV effects (\ref{a12}%
, \ref{a74}) is used , this rephasing factor $\varphi _{B}^{\prime }$ changes nothing
because it is equal to one 
\begin{eqnarray}
\rho _{B} &=&-\exp j4\beta \text{,}  \label{a88} \\
\varphi _{B}^{\prime } &=&1\text{.}  \label{a89}
\end{eqnarray}

If CPV is gravity induced, we replace $\left| B_{H}\right\rangle $ and $%
\left| B_{L}\right\rangle $ with $\left| B_{H}^{\oplus }\right\rangle $ and $%
\left| B_{L}^{\oplus }\right\rangle $ given by (\ref{a70}, \ref{a72}), and
the very same observable $\rho _{B}\varphi _{B}^{\prime }$ is the product of the following factors 
\begin{eqnarray}
\rho _{B}^{\oplus } &=&-1+O\left[ 10^{-6}\right] \text{,}  \label{a90} \\
\varphi _{B}^{\prime \oplus } &=&\frac{1+j\chi }{1-j\chi }=\exp \left(
2j\arctan \chi \right) \text{.}  \label{a91}
\end{eqnarray}
We conclude that, within the framework of GICPV, the measurement of the
phase-convention-independent observable $\rho _{B}\varphi _{B}^{\prime }$ on
Earth gives 
\begin{equation}
\rho _{B}^{\oplus }\varphi _{B}^{\prime \oplus }=-\exp \left( 2j\arctan \chi
\right) \text{,}  \label{a92}
\end{equation}
although if the measurement of the very same phase-convention-independent
observable $\rho _{B}\varphi _{B}^{\prime }$ is interpreted within the usual
SM-CKM framework it defines $\beta $ as 
\begin{equation}
\rho _{B}\varphi _{B}^{\prime }=-\exp j4\beta \text{.}  \label{a93}
\end{equation}
The conclusion of this $\rho _{B}$ {\it gedanken experiment} measurement, with two
frameworks of interpretation, is that $\arctan \chi =2\beta $ and 
$
\sin 2\beta =\sin \left[ \arctan \left( 0.77\right) \right] =0.61
$.

The
measurement of the angle $\beta $ is still one of the major subjects at the
forefront of the studies related to the physics of the SM and beyond.

\section{Conclusion}

On the basis of highly accurate predictions of both $\varepsilon$ 
(\ref{a41}, \ref{a4522}) 
and $\varepsilon
^{\prime }$ (\ref{a57}), and of a relevant prediction of $\beta $ (\ref{a184}), we can state
that GICPV offers a pertinent framework to interpret $K^{0}/%
\overline{K}^{0}$ and $B^{0}/\overline{B}^{0}$ experiments dedicated to CPV
and that the CKM matrix should be considered free from any CPV phase far from
any massive object.

Gravity induced Charge-Parity violation not only explain ({\it i}, {\it ii} and {\it iii}) CPV
effects, and predict the associated observables, but it also renews, in depth, 
the baryons
asymmetry ({\it iv}) cosmological issue.

The previous calculations on the impact of Earth gravity on neutral mesons
oscillations can be extended to $D^{0}/\overline{D}^{0}$ $\sim \left(c%
\overline{u}\right) /\left( \overline{c}u\right) $ and $B_{s}^{0}/\overline
{B_{s}}^{0}\sim \left( s\overline{b}\right) /\left( \overline{s}b\right) $.
The framework of analysis of the experimental data on $D^{0}/\overline{D}%
^{0} $ and $B_{s}^{0}/\overline{B_{s}}^{0}$ is similar to the methods
presented in section 5, 6 and 7. The parameters $m_{D}g\hbar /\delta
m_{D}^{2}c^{3}$ and $m_{B_{s}}g\hbar /\delta m_{B_{s}}^{2}c^{3}$ for both
mesons systems are very small so a type ({\it i}) indirect violation will
be extremely difficult to observe. However type ({\it ii}) and type ({\it iii%
}) CPV can be analyzed on the basis of a clear definition of phase-convention-independent 
observables, similar to those identified in
section 6 and 7, and will be considered in a forthcoming study.

In any environment where a flavored neutral mesons $\left| M\right\rangle $,
with mass $m$, mass spliting $\delta m$ and Compton wavelength $\lambda _{C}$, 
experiences a gravity ${\bf g}$, i.e. in any curved space-time
environment, a CP eigenstates mixing with amplitude 
$j\left( m/\delta m\right) ^{2}\left| {\bf g}\right| \lambda _{C}/c^{2}$ will be observed.
The first factor $m/\delta m$ is associated with electroweak and strong
interactions, the second one is the product of a (wave)length, an
acceleration and $c$, quantities related to space-time geometry rather
than to electroweak or strong interactions. For a massive spherical object, 
with radius $R_{0}$ and Schwarzschild radius $R_{S}$, the mixing amplitude is given by
$j\left( m/\delta m\right) ^{2}\left( R_{S}/R_{0}\right)(\lambda _{C}/R_{0})$.
The proportionality to $\left| 
{\bf g}\right| $ or $ R_{S}$ indicates that this new CPV mechanism allows to set up
cosmological evolution models predicting the strong asymmetry between the
abundance of matter and the abundance anti-matter in our present universe 
\cite{6}.

Beside the problem of early baryogenesis, neutrinos oscillations near a
spherical massive object might be revisited to explore the impact of the
interplay between gravity and mixing.

The type ({\it i}) CPV observed with $K^{0}/\overline{K}^{0}$ stems from a
gravity induced interplay between vertical quarks zitterbewegung
oscillations at the velocity of light on the one hand and the strangeness
oscillations $\left( \Delta S=2\right) $ on the other hand.

The type ({\it ii}) small CPV observed with $K^{0}/\overline{K}^{0}$ is
associated with the SM-CKM, CPT invariant, interpretation of a GICPV 
and is elucidated through a careful analysis of the rephasing
invariance of the observable $\eta _{00}$.

The large type ({\it iii}) CPV observed with $B^{0}/\overline{B}^{0}$ is
associated with the SM-CKM, CPT invariant, interpretation of a GICPV, displaying a 
very small modulus and a significant phase, and is elucidated through a careful 
analysis of the rephasing
invariance of the observable $\beta$.

When the mesons are considered stables, the evolution is unitary and there
is no T violation at the fundamental level. The observed T violation stems 
from very the 
large phase space offered to the final states of forward transitions  $M\rightarrow f$ 
so that  backward
transitions $f\rightarrow M$ can not be observed. The
amplitudes $w_{f}$  in equation (\ref{a1}) can be considered within the framework 
of the WW approximation \cite{13}
as irreversible decays in equation (\ref{a2}).

The very large type ({\it iv}) CPV observed in our universe remains an open 
issue within the SM-CKM framework
of interpretation, whereas GICPV displays the potential to
set up cosmological evolution models in agreement with the present state of
our universe.

This set of new results is obtained without any speculative assumption on new
fields, and without any free parameters adjustment. From this convergence of
results, we can conclude that a CKM matrix free of CPV phase is to be
considered as the core of the SM in a flat Lorentzian environment and
Earth's gravity is the sole source of $\ \varepsilon $, $\varepsilon
^{\prime }$ and $\beta $ CPV  effects observed
 in $K^{0}/\overline{K}^{0}$ and $B^{0}/ 
\overline{B}^{0}$ dedicated experiments.


\begin{thebibliography}{99}

\bibitem{1}  J. H. Christenson and J. W. Cronin and V. L. Fitch and R.
Turlay. 1964 Evidence for the $2\pi $ decay of the $K_{2}^{0}$ meson. {\it %
Phys. Rev. Lett.} {\bf 13}, 138-140. (doi:10.1103/PhysRevLett.13.138)

\bibitem{2}  M. Kobayashi and T. Maskawa. 1973 CP-violation in the
renormalizable theory of weak interaction. {\it Progress of Theoretical
Physics} {\bf 49}, 652-657. (doi:10.1143/PTP.49.652)

\bibitem{3}  N. Cabibbo. 1963 Unitary symmetry and leptonic decays. {\it %
Phys. Rev. Lett.} {\bf 10}, 531-533.(doi:10.1103/PhysRevLett.10.531)

\bibitem{4}  T. D. Lee. 1981 {\it Particle physics and introduction to field
theory}. New York, USA: Harwood Academic.

\bibitem{5}  P. D. Group, Review of particle physics, Phys. Rev. D {\bf 110}
030001 (2024). (doi:10.1103/PhysRevD.110.030001)

\bibitem{6} A. D. Sakharov, Violation of CP invariance, C asymmetry, and
baryon asymmetry of the universe, Pisma Zh. Eksp. Teor. Fiz. 5, {\bf 32}
(1967). (doi:10.1070/PU1991v034n05ABEH002497, 10.3367/UFNr.0161.199105h.0061)

\bibitem{7}  E. Fischbach, Test of general relativity at the quantum level,
in {\it Proceedings of the NATO Advanced Study Institute on Cosmology and
Gravitation}, edited by P. Bergmann and V. de Sabbata, NATO Scientific
Affairs Division (Plenum Press, New york and London, 1980) pp. 359-373. 


\bibitem{8}  G. Chardin and J-M. Rax, CP violation. a matter of
(anti)gravity?, Phys. Lett. B {\bf 282}, 256 (1992). (doi:10.1016/0370-2693(92)90510-B)

\bibitem{9}  J. Bjorken and S. Drell, {\it Relativistic Quantum Mechanics}
(McGraw-Hill, New york, 1964).

\bibitem{10}  T. D. Lee, R. Oehme, and C. N. Yang, Remarks on possible
noninvariance under time reversal and charge conjugation, Phys. Rev. {\bf 106%
}, 340 (1957). (doi:10.1103/PhysRev.106.340)

\bibitem{11}  T. T. Wu and C. N. Yang, Phenomenological analysis of violation
of CP invariance in decay of $K^{0}$ and $\overline{K}^{0}$, Phys. Rev.
Lett. {\bf 13}, 380 (1964). (doi:10.1103/PhysRevLett.13.380)

\bibitem{12}  J-M. Rax, Zitterbewegung CP violation in a Schwarzschild
spacetime, arXiv 2403.07970 (03-2024). (doi:10.48550/arXiv.2403.07970)

\bibitem{13}  V. Weisskopf and E. Wigner, Berechnung der naturlichen
linienbreite auf grund der diracschen lichttheorie, Zeitschrift fur Physik 
{\bf 63}, 54 (1930). (doi:10.1007/BF01336768)

\bibitem{14} J-M. Rax, Gravity induced CP violation, arXiv 2405.17317
(05-2024). (doi:10.48550/arXiv.2405.17317)

\bibitem{15}  J. Bell and J. Steinberger, in {\it Proceedings of the Oxford
International Conference on Elementary Particles 1965}, edited by R. G.
Moorhouse, A. E. Taylor, and T. R. Walsh (Rutherford High Energy Laboratory,
1966) p. 195. 

\bibitem{16}  H. Burkhardt et al. (NA31), First evidence for direct CP
violation, Phys. Lett. B {\bf 206}, 169 (1988). (doi:10.1016/0370-2693(88)91282-8)

\bibitem{17}  V. Fanti et al. (NA48), A new measurement of direct CP
violation in two pion decays of the neutral kaon, Phys. Lett. B {\bf 465},
335 (1999). (doi:10.1016/S0370-2693(99)01030-8)

\bibitem{18}  A. Alavi-Harati et al. (KteV), Observation of direct CP
violation in $K_{S,L}\rightarrow \pi \pi $ decays, Phys. Rev. Lett. {\bf 83}%
, 22 (1999). (doi:10.1103/PhysRevLett.83.22)

\bibitem{19}  B. Aubert et al. (BaBar), Observation of CP violation in the $%
B_{0}$ meson system, Phys. Rev. Lett. {\bf 87}, 091801 (2001). 
(doi:10.1103/PhysRevLett.87.091801)

\bibitem{20}  K. Abe et al. (Belle), Observation of large CP violation in
the neutral $B$ meson system, Phys. Rev. Lett. {\bf 87}, 091802 (2001).
(doi:10.1103/PhysRevLett.87.091802)

\bibitem{21}  B. Aubert et al. (BaBar), Improved measurement of CP violation
in neutral $B$ decays to $c\overline{c}s$, Phys. Rev. Lett. {\bf 99}, 171803
(2007). (doi:10.1103/PhysRevLett.99.171803)

\bibitem{22}  I. Adachi et al. (Belle), Precise measurement of the CP
violation parameter, Phys. Rev. Lett. {\bf 108}, 171802 (2012). (doi:10.1103/PhysRevLett.108.171802)

\bibitem{23}  B. Aubert et al. (BaBar), Measurement of time-dependent
asymmetry in decays, Phys. Rev. D {\bf 79}, 072009 (2009). (doi:10.1103/PhysRevD.79.072009)

\end{thebibliography}
\end{document}